\documentclass[10pt]{iopart}
\usepackage{epsf}

\usepackage{bm}

\begin{document}
\title[Competition between a uniform source and a local trap in a stochastic system]{ Competition between homogeneous and local processes
 in a diffusive many-body system }

 \author{Mauro Mobilia\footnote[1]{email:mmobilia@vt.edu}}
 \address{
 Center for BioDynamics,
  Center for Polymer Studies and Department of Physics, Boston University,
  Boston, MA, 02215, USA 

    \vspace{0.3cm}
  
  Center for Stochastic Processes in Science and Engineering, Department of Physics, Virginia Polytechnic Institute and State University, Blacksburg, VA 24061-0435, USA  \footnote{Present address}
  }

\begin{abstract}
We consider a stochastic many-body system where a source refills uniformly the empty sites of a hypercubic
lattice, on which each particle is allowed to  jump (symmetrically) onto neighboring vacant sites.
In addition, there is a local {\it trap}, in competition with the injection reaction,  which 
perturbs the dynamics by removing particles from the system. In dimensions $d=1, 2$ and $3$, for an ``imperfect'' and a ``perfect'' trap,  the spatiotemporal effect of  the local perturbation of the dynamics is investigated by computing the exact
concentration of particles in the system and it is shown  that the density profile exhibits a depletion (in one and two dimensions) which properties depend on the space dimension. The exact reactive (fluctuating) steady state and the long-time behavior of the concentration of particles are explicitly computed.
\end{abstract}
\pacs{02.50.-r, 02.50.Ey, 05.50.+q, 05.40.Jc}

\maketitle
\date{\today}

\section{Introduction}
In the last decades much effort has been devoted to the
study of nonequilibrium systems, both theoretically
(see e.g. \cite{Privman,Schutzrev,DPAC} and references therein) and experimentally (see e.g.
\cite{TMMC,MX,Phototrap,Photobleach2D} and references therein). 
Despite these efforts,
a comprehensive theoretical framework is still lacking: As yet, there is no
equivalent of ensemble theory for nonequilibrium systems. Consequently,
in nonequilibrium systems one has to explicitly impose the dynamics in order to be able 
to compute observable quantities. In this context the stochastic dynamics 
is generally described by a master equation, and most progress in the
field is made by studying paradigmatic models \cite{Privman}. In this
framework, \emph{exact} solutions of simple models are scarce, but very
precious, since they can serve as benchmarks for approximate and/or
numerical schemes and shed light on general properties of whole classes of
related models. Not surprisingly, \emph{nontrivial} solutions are almost
entirely restricted to one dimension (1D; see e.g. \cite{Privman,Schutzrev}
) and have focused mainly on spatially homogeneous models. 
One should however mention that the authors of Ref. \cite{Rittenberg} were able 
to obtain the exact steady-state of a one-dimensional coagulation-diffusion model with local 
particle input, while the author of Ref. \cite{benA} computed the non-trivial stationary concentration
 of particles in a one-dimensional system of reversible diffusion-limited coalescence 
in the presence of a (perfect) static trap. Also, recently the inhomogeneous version of the voter model, and its mapping to the monomer-monomer catalytic reaction have been solved in arbitrary dimensions \cite{IVM,MI}.

Actually, in many realistic
situations, diffusion of particles (describing {\it e.g.}  ions or  excitons \cite{Privman,TMMC,MX}) 
 often takes place at interfaces of different phases (inhomogeneous substrates) and thus the understanding
 of the spatiotemporal  effects of impurities, defects, and inhomogeneities 
  is highly desirable. To gain some further insight in these
 issues, the analysis of simple models seems to be the most natural and
promising approach. Motivated by these considerations, and complementary to 
recent studies on inhomogeneous reaction-diffusion systems (see e.g. \cite{Rittenberg,benA}), we specifically
investigate, through {\it exact methods} and in {\it arbitrary dimensions}, the properties of an inhomogeneous  diffusive stochastic many-body system  where particles are injected by a uniform source which refills empty sites of a lattice. 
Once on the lattice the particles can 
jump symmetrically onto their empty neighboring sites. We also consider that at the origin of the lattice 
there is {\it trap} that can remove particles from the system.
The goal here is to carefully analyze the effect of the competition between the uniform source and the local trapping reaction in various dimensions and to study the spatiotemporal concentration profile of particles around the trap. It is therefore shown that the interplay of these competing (homogeneous vs local) reactions results
in nontrivial stationary states and affects also the time behavior of the system. 
In addition, as there have been recently various experimental works \cite{Phototrap,Photobleach2D}
devoted to the study of the properties of such a model in the absence of the source, it seems that our 
study could also have some experimental implications as well. 

This work is organized as follows: In the next
section we give a precise  mathematical definition of the model under
consideration, the basic equation and an overview of the main results.
In Section 3 we  set-up the formalism to compute the dynamical concentration of particles.
In Section 4, we especially focus on the 1D infinite chain and obtain
explicitly the stationary and long-time behavior of the concentration.
In Section 5, we focus on the two-dimensional infinite system, while the section 6 deals with the 
three-dimensional case (in the thermodynamic limit).
In Section 7, we study the global effect of the trap and the source for infinite systems in 1D, 2D and 3D.
In Appendix A, we briefly review the results already obtained (through other methods \cite{Ben,Weiss,Taitelbaum2,Taitelbaum}) for 
the diffusive model in the presence of the trap but in the {\it absence} of the source (and in the thermodynamic limit).
In Appendix B, one extends the discussion to the case of {\it periodic boundary conditions} on finite lattices and especially focuses on the concentration of particles on a ring (1D) of finite size. 
\begin{center}
\begin{table}[t]
\caption{The three elementary processes under consideration.
As explained in the text, the hopping and injection processes are homogeneous, whereas
the trapping reaction is localized at site ``${\bm 0}$". }
\begin{indented}
\item[]
\begin{tabular}{@{}ll}
\br
Reactions & Processes \\
\mr
 $A \emptyset \stackrel{D}{\longleftrightarrow}\emptyset A$    & Symmetric hopping with exclusion\\ 
$ \emptyset  \stackrel{J}{\longrightarrow}A$  & Homogeneous source   refilling empty sites   \\
$A   \stackrel{{\cal T}}{\longrightarrow} \emptyset$  & Trapping of particles at site ``${\bm 0}$''\\
\br
\end{tabular}
\end{indented}
\end{table}
\end{center}

\section{The Model and outline of the results}

\begin{figure}
\begin{center}
\epsfbox{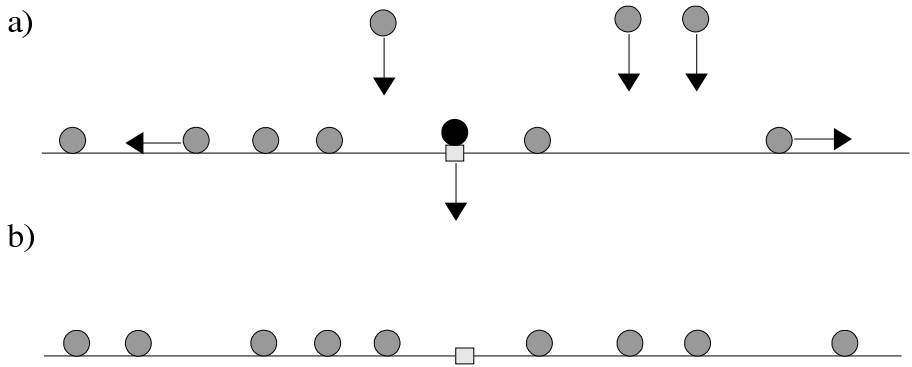}
\end{center}
\caption{\label{Fig1}Illustration of the processes occurring in the one-dimensional version of the model. a) Particles can jump to neighboring empty site (with rate $D$) and can be injected on empty sites by a source of intensity $J$. The site at the center of the lattice is the ``trap'' and the particles (indicated in black) landing 
on this site can be removed with a rate ${\cal T}$. b) Configuration of the system after all the moves sketched in a) have been performed.}
\end{figure}

In dimensions $d=1, 2$ and $3$, we consider a stochastic model of
interacting particles  on a hypercubic lattice  of size $(2L+1)^{d}$.
 Each site, that can {\it at most be occupied by one
 particle},
is labelled by a vector ${\bm x}$ of components
 $-L\leq x_{i}\leq L, \, i=1,\dots, d$. On this lattice, particles denoted $A$
 can jump (subject to the site restriction) to a neighboring empty site (denoted by the symbol $\emptyset$) with a rate $D$ and an uniform source refills the empty sites of the lattice by injecting particles with rate $J$. In addition to these homogeneous processes, we consider that
  there is a {\it trap} at the origin of the lattice such that a particle landing on this ``poisoned'' site can be trapped and removed from the system with a rate ${\cal T}$. These processes are summarized in the Table 1 and illustrated in the Figure 1. 
  
 The competition between the homogeneous process refilling the system and 
 the local one trapping the particles is studied by investigating the properties of the 
 concentration of particles. The equation of motion for such a quantity is obtained from the 
 master equation and reads :
  
 \begin{eqnarray}
\label{nxdot}
\frac{d c_{\bm x}(t)}{dt}=D \ \Delta_{\bm x}c_{\bm x}(t) + J[1-c_{\bm x}(t)] -{\cal T}c_{\bm 0}(t) \delta_{{\bm x},{\bm 0}}.
\end{eqnarray} 
Here $\Delta_{\bm x}$ denotes the discrete Laplace operator:
$\Delta_{\bm x}c_{\bm x}(t)\equiv
 - 2d c_{\bm x}(t) +\sum_{{\bm x'} } c_{\bm x'}(t)$, where the sum on right-hand side (r.h.s.) 
 runs over the $2d$ nearest neighbors ${\bm x'}$
of site ${\bm x}$. It should be noted that the equation (\ref{nxdot}) is both valid for infinite systems ($L\to \infty$) and for periodic lattices of arbitrary size.
In Sections 3 to 7, for the sake of technical simplicity, Eq. (\ref{nxdot}) will be studied in the thermodynamic limit ($L\to \infty$). The case of periodic boundary condition, somewhat technically more involved, is briefly addressed in Appendix B.

At this point, some comments on the diffusion-like Eq. (\ref{nxdot}) are in order.
We first notice that the effect of the trap appears through the inhomogeneous term $-{\cal T}c_{\bm 0}(t) \delta_{{\bm x},{\bm 0}}$ which is non zero only at the origin of the lattice and is proportional to the concentration.
In addition to the inhomogeneity, the equation is not merely diffusive since there is a term accounting for the
injection of particles by the source. Because of the site restriction the system allows at most one particle at each site. Consequently the source term refills only empty sites and contributes with a term $J[1-c_{\bm x}(t)]$ (and not simply $J$).
One can also mention that systems with site restrictions are usually not solvable in arbitrary dimensions: when available, most of the exact solutions are obtained in one-dimension 
(see e.g. \cite{Privman,Schutzrev} and references therein). It is also worth noting that the authors of Refs. \cite{Rittenberg,benA} were able to solve some spatially inhomogeneous versions of the coagulation-diffusion stochastic model in one dimension by using the so-called IPDF method (see {\it e.g.} \cite{Privman} and references therein).
 
 Remarkably, for the model under consideration, the equation (\ref{nxdot}) for the concentration is linear (but inhomogeneous) and thus solvable.
Obviously, in the absence of the source term, the system reduces to a purely diffusive model in the presence of a trap and, because the hopping process is symmetric, the volume exclusion does not affect the concentration of the 
particles [see Eq.(\ref{nxdot})]. In this case one recovers the results obtained in previous studies based on non-interacting random walks  \cite{Phototrap,Photobleach2D,Ben,Weiss,
Taitelbaum2,Taitelbaum}, as it is briefly reviewed (for a discrete lattice) in Appendix A. Also, it is clear that in the absence of the trapping reaction (${\cal T}=0$ and $J>0$), the system is eventually completely occupied by the particles and (for homogeneous initial conditions), the approach toward the steady state is purely exponential in all dimensions: $c_{\bm x}(t)-1\sim e^{-J t}$ [see Eq.(\ref{sol})]. In the presence of the source, the role of the site restriction {\it does effectively matter}. In fact, if one would relax such a constraint [and would have $J$ instead of $J(1-c_{\bm x}(t))$ in Eq.(\ref{nxdot})], the concentration of particles would simply increase linearly with the time and there would be no effective competition with the trapping reaction. In this sense, the many-body character of the system under consideration is embodied in Eq.(\ref{nxdot}) by the term $J(1-c_{\bm x}(t))$: as the source injects particles only on empty sites, the interplay with the trap drives the system toward a non-trivial genuine nonequilibrium reactive (fluctuating) steady state which properties explicitly depend on the dimension of the system. Of course, the most interesting situation arises when the homogeneous injection rate is small (but not zero) and is thus efficiently balanced by the trapping rate.
In fact, as we show in the next sections, the effect of the competition between the local/homogeneous reactions
(on infinite systems) results in a perturbation of the uniform static density profile $\delta c_{\bm x}(\infty)\equiv 1-c_{\bm x}(\infty)$ which reads ($x\equiv |{\bm x}|$)

\begin{eqnarray}
\label{results}
\delta c_{\bm x}(\infty)
=\cases{ 
A_{1}\, e^{-\xi x} &;   in 1D \\  
\frac{2K_{0}(x\sqrt{J/D})}{\ln{(D/J)}} &;   in 2D \\ 
A_{3}\,e^{-x\sqrt{J/D}}/x
 &;  in 3D,\\}
\end{eqnarray}
where $\xi\equiv 2\ln{\left(\frac{\sqrt{J}+\sqrt{J+4D}}{2\sqrt{D}}
\right)}<\sqrt{J/D}$, $K_0(z)$ is the Bessel function of third kind \cite{Abramowitz} and $A_{1,3}$ are amplitudes which values are given by Eqs 
(\ref{cstat1d}) and (\ref{cstat3d}). It is worth mentioning that the 1D and 3D results quoted in Eq.(\ref{results})
are valid for any value of the parameters. However, while the 1D expression is the exact solution of the stationary difference equation (\ref{nxdot}), the 3D result has been obtained from a suitable continuum reformulation of Eq.(\ref{nxdot}).
On the other hand, the above 2D expression of $\delta c_{{\bm x}}(\infty)$ has been obtained from Eq.(\ref{nxdot}) in the limit where $J$ is small.

 We also show that the time evolution of the concentration is affected by the competition between the source and the trap and depends on the dimensionality of system. In fact, it is found (on infinite systems) that the time-dependence is not purely exponential and depends on the initial state of the lattice:
\begin{eqnarray}
\label{time_evol}
c_{\bm x}(t) -c_{\bm x}(\infty) \propto
\cases{ 
e^{-Jt} t^{-\nu_{1}}&;    in 1D \\ 
e^{-Jt} (\ln{t})^{-\eta} &;  in 2D \\
e^{-Jt} t^{-\nu_{3}}
 &; in 3D,\\}
\end{eqnarray}
where the subdominant time-dependent prefactors have the exponents $\nu_{1}=\nu_{3}=3/2$ and $\eta=2$ for an initially fully occupied lattice and $\nu_{1}=\nu_{3}=1/2, \eta=1$ for other (homogeneous) initial conditions.
All these results are derived and discussed in detail in the remainder of this work.
\section{General set-up}

Using the properties of the Bessel functions of first kind 
$I_{n}(z)\equiv \int_{0}^{\pi} \frac{dq}{\pi}e^{z \cos{q}} \cos{nq}$,
\cite{Abramowitz} and extending the techniques devised and briefly presented
in Ref. \cite{IVM} (see also \cite{MI}), one derives from equation (\ref{nxdot}) 
the following exact and self-consistent relationship obeyed by the concentration in the thermodynamic limit ($L\to \infty$):
\begin{eqnarray}
\label{sol}
 c_{\bm x}(t) &=& 1+ \sum_{\bm y} (c_{\bm y}(0)-1) e^{-Jt}\prod_{i=1}^{d}
 \left[e^{-2Dt} I_{x_i-y_i}(2Dt)\right] \nonumber\\
 &-&{\cal T} \int_{0}^{t} dt' c_{\bm 0}(t-t') \, e^{-Jt'} \prod_{i=1}^{d}
 \left[e^{-2Dt'} I_{x_i}(2Dt')\right], 
\end{eqnarray}
where the sum on the r.h.s takes care of the initial condition and the convolution term
accounts for the inhomogeneity. Of course, this expression is valid both in the absence
 ($J=0$) and in the presence ($J>0$) of the source. It is clear from this expression that in 
 the absence of the trapping reaction ({\it i.e.} if ${\cal T}=0$), at long time 
 the lattice is fully occupied: $c_{\bm x}(\infty)=1$.
 Obviously, the situation is much more interesting and subtle when there is a 
 competition between the trap and the source.
It is also important to notice that the explicit determination of the density of particles still
 requires the solution of the self-consistent integral equation (\ref{sol}),
 where the expression of the concentration appears both on
 the right (r.h.s) and left hand-side (l.h.s).
This is achieved by using Laplace transform techniques and the integral
 representation of the modified Bessel functions.
In fact, we introduce the quantity
${\cal I}_{\bm x}$ that is essentially the Laplace transform of a product of Bessel functions (multiplied by an exponential factor):
\begin{eqnarray} 
\label{^Ix}
 {\cal I}_{\bm x}(s,D,J)\equiv \int_{0}^{\infty} dt \, e^{-(s+J)t} \; 
 \prod_{i=1}^{d}\left\{e^{-2Dt}
 I_{x_i}(2Dt)
 \right\}.
\end{eqnarray}
From the  integral representation of the modified Bessel functions, one can check that 
the quantities ${\cal I}_{\bm x}(s,D,J)$, hereafter simply  denoted  ${\cal I}_{\bm x}(s)$,
 can be rewritten as the so-called Watson integrals 
(or cubic lattice Green-functions):
\begin{eqnarray} 
\label{Wat}
 {\cal I}_{\bm x}(s)=\int_{-\pi}^{\pi} \dots \int_{-\pi}^{\pi} \frac{d^{d}{\bm q}}
 {(2\pi)^{d}} \frac{e^{i {\bm q}.{\bm
 x}}}{s+J+2D\sum_{i=1}^{d}(1-\cos{q_i})},\nonumber\\
\end{eqnarray}
where we have introduced a $d$-dimensional vector ${\bm q}=(q_1, \dots, q_d)$.
For further convenience we introduce  the norm of the vector ${\bm x}$, that
is denoted $ x\equiv |{\bm x}|\equiv \sqrt{\sum_{i=1}^{d}x_{i}^{2}}\geq 0$, 
and the parameter $u\equiv x\sqrt{J/D}$. Also, as in arbitrary dimensions the concentration of particles 
 depends spatially only on the distance $x$ to the origin [see e.g. Eq. (\ref{sol})], 
 hereafter it is simply denoted $c_{x}(t)$
 (instead of $c_{\bm x}(t)$).

We now focus on the  site ${\bm x}={\bm 0}$, and  specifically consider
that initially the system is randomly occupied with an homogeneous concentration
$c_{\bm x}(0)=\rho_0$ of particles.
Using the convolution theorem and the identity $\sum_{n=-\infty}^{+\infty}
I_{n}(z)=e^{z}$ \cite{Abramowitz}, from Eq. (\ref{sol}) we get the Laplace transform 
of the concentration of particles at the origin:
\begin{eqnarray} 
\label{n0(s)}
 \hat{c}_{0}(s)&\equiv& \int_{0}^{\infty} dt \,  e^{-st} \, c_{0}(t)=
  \frac{1}{[1+{\cal T} {\cal I}_{\bm 0}(s)]} \left(
\frac{1}{s}-\frac{1-\rho_0}{s+J}
\right).
\end{eqnarray}
Plugging back the result (\ref{n0(s)}) into the Laplace transform of Equation (\ref{sol}), we obtain:
\begin{eqnarray} 
\label{nx(s)}
 \hat{c}_{x}(s)&\equiv& \int_{0}^{\infty} dt \, e^{-st}c_{x}(t)=
  \frac{1+{\cal T} [ {\cal I}_{\bm 0}(s) -{\cal I}_{\bm x}(s) ]}{1+{\cal T}
  {\cal I}_{\bm 0}(s)}\left[\frac{1}{s} -\frac{1-\rho_0}{s+J}\right].
\end{eqnarray}

Again, the expressions (\ref{n0(s)}) and (\ref{nx(s)}) are valid for any strength $J\geq 0$
of the source.
Using the general properties of Laplace transforms
\cite{Abramowitz}, the concentration for a system with initially
$c_{x}(0)=\rho_0$ is simply related to the concentration obtained from the
case where $\rho_0=1$ and $J=0$. In fact, it readily follows from (\ref{n0(s)}) and (\ref{nx(s)})
that 
\begin{eqnarray} 
\label{CI}
  {c}_{ x}(t)|_{_{J\geq 0; 0\leq \rho_0\leq 1}}=
 {c}_{x}(t)|_{_{ J\geq 0; \rho_0=1}} 
-(1-\rho_0)e^{-Jt} \left[{c}_{x}(t)|_{_{J=0; \rho_0=1}}\right] .
\end{eqnarray}
The long-time behavior of the concentration of particles is therefore
obtained from the analysis of (\ref{n0(s)}) and (\ref{nx(s)}). 
This study is the scope of the the subsequent sections where due 
attention is paid to the competition between the homogeneous/local reactions and to 
the effect of the dimensionality on the spatiotemporal properties of the system.

\section{One-dimensional concentration in the presence of the source and the trap}

In this section we compute explicitly the stationary concentration as well as 
its long-time behavior on an infinite chain (1D).
    
In one dimension, the quantity
${\cal I}_{x}(s)$ can be computed explicitly and one has ($x>0$, see Ref. \cite{Abramowitz}):
\begin{eqnarray} 
\label{Ix1d}
  {\cal I}_{x}(s)=\frac{1}{\sqrt{(s+J)(s+J+4D)}}
   \left\{\frac{\sqrt{s+J+4D}-\sqrt{s+J}}{2\sqrt{D}}\right\}^{2x}
\end{eqnarray}

It is clear from this expression that, as long as $J>0$, ${\cal I}_{ x}(s)$ is
 a well defined quantity, even in the  $s\rightarrow 0$ limit.
However, when $J=0$, in the regime  $s\rightarrow 0$, ${\cal I}_{ x}(s)$ 
diverges as $s^{-1/2}$. Because of this fact, we can already anticipate an
 exponential dynamics in the presence of the source and an algebraic one
in absence of the latter (see Appendix A). 
In this section, we focus on the case where the source is non-zero.

Clearly, it follows from (\ref{nx(s)}), that the stationary concentration reads (for arbitrary initial conditions):
\begin{eqnarray} 
\label{cstat}
 c_{x}(\infty)=1-\frac{{\cal T} {\cal I}_{x}(0)}{1+{\cal T} {\cal I}_{0}(0)}.
\end{eqnarray}

In fact, the formula (\ref{cstat}) holds in any 
dimensions (when $J>0$): 
The dimension $d$  enters the expression (\ref{cstat}) of 
 $c_{x}(\infty)$ through the value of the integrals (\ref{Wat}) appearing 
in it.
 In one dimension, for $J>0$, with (\ref{cstat}) and  (\ref{Ix1d}), we find that the 
stationary concentration approaches the value $1$ exponentially with the distance to 
the origin:
\begin{eqnarray} 
\label{cstat1d}
 c_{x}(\infty)=1-\frac{{\cal T} \; e^{-\xi x}}{{\cal T} +\sqrt{J(J+4D)}},
\end{eqnarray}
where one has introduced $\xi\equiv 2\ln{\left(\frac{\sqrt{J}+
\sqrt{J+4D}}{2\sqrt{D}}\right)}>0$. 
One notices that $\xi<\sqrt{J/D}$ and when $J\ll 1$ one has $\xi=\sqrt{J/D}\, 
\left[1 - J/24D +{\cal O}((J/D)^{2})\right]$. For a perfect trap, when ${\cal T}
\rightarrow \infty$ (with $D$ and $J$ finite), the formula
(\ref{cstat1d}) gives: $c_{x}(\infty)=1-e^{-\xi x}$ and $c_{0}(\infty)=0$.
These results indicate that around the origin the competition between the source 
and the trap generates a depletion zone of length $\ell_{1D}= 2\xi^{-1}$.
Interestingly, in the reaction-diffusion system $AA \leftrightarrow A$ it was also shown 
that a (perfect) trap also depletes the concentration profile with an effect decaying exponentially with the distance to the inhomogeneity \cite{benA}.
Here, the situation is particularly interesting in the limit of a {\it weak source}, when
$J/{\cal T} \ll 1$ and $J/D \ll 1$, but $J>0$. In this case, the competition between the 
localized trap and the uniform source is really effective and one shall distinguish two regimes ($x>0$):
 (i) when $0 \leq u\equiv x(J/D)^{1/2}\ll 1$, the concentration of particles increases linearly with the distance
 from the trap and with an amplitude proportional to $\sqrt{J}$ : $c_x(\infty)=u $; (ii) far away from the trap, when $u$ is large enough, the effect of the local perturbation disappear exponentially with the distance to the origin: $\delta c_x(\infty)= e^{-u}$. When $J$ is small, the size of the depletion region is $\ell_{1D}\sim (D/J)^{1/2}$.
To describe further the static properties of
 this zone we can also introduce a ``$\theta_s$-distance'' ($x_{\theta_{1D}}(\infty)$), which is the {\it static} distance from the trap to the point where 
the stationary  concentration of $A$ particles reaches the specific fraction
$\theta_s$  of unity ({\it i.e.} of the stationary concentration in the bulk).
Requiring that  $c_{x}(\infty)= \theta_s $ on the l.h.s of 
 Eq. (\ref{cstat1d}), we obtain ($\sqrt{D/J} \gg 1$ and $\sqrt{{\cal T}/J} \gg 1$):
\begin{eqnarray}
\label{xthetas1d}
x_{\theta_{1D}}(\infty)=
\theta_{s}\sqrt{\frac{D}{J}} -\frac{2D}{{\cal T}},
\end{eqnarray}
which means that $x_{\theta_{1D}}$ varies linearly with $\theta_s$.

We now compute the asymptotic approach  to the stationary concentration and start
 with a system initially completely occupied ($\rho_0=1$).
By noticing that
$c_{x}(t) =1- \int_{0}^{t} dt' e^{-Jt'} 
{\cal L}^{-1} \left[
\frac{{\cal T} [{\cal I}_{x}(s)|_{_{J=0}}]}
{1+{\cal T} [{\cal I}_{ 0}(s)|_{_{J=0}}]}\right](t')$, where ${\cal L}^{-1}[\dots]$ denotes the usual inverse-Laplace-transform, the long-time behavior is obtained from the
 Laplace-inversion of the
$s\rightarrow 0$ expansion of the integrand of such an expression and one thus gets  ($t\to \infty$):
\begin{eqnarray*} 
\label{nx1CII}
 c_{0}(t)-c_{0}(\infty)&\simeq & \frac{D}{{\cal T}}\,\frac{e^{-Jt}}{Jt\,(\pi D t)^{1/2}},
\end{eqnarray*}
and when $0<x<\infty$
\begin{eqnarray} 
\label{nx1d}
 c_{x}(t)-c_{x}(\infty)&\simeq&  
\frac{x\,e^{-Jt-x^2/4Dt}}{Jt\,(\pi D t)^{1/2}},
\end{eqnarray}
where $c_0(\infty)$ and $c_x(\infty)$ are given by (\ref{cstat1d}).
Together with the results of Appendix A [see Eq. (\ref{nx(t)1dss})], we
can  immediately obtain the long-time behavior of the concentration
for  an initially random system, where $0\leq \rho_0 \leq
1$. In fact, according to 
(\ref{CI}) we obtain ($0\leq \rho_0 <1$ and  $t\to \infty$): 
\begin{eqnarray} 
\label{nx1CI}
 c_{0}(t)-c_{0}(\infty)&\simeq & -\frac{2(1-\rho_0)}{{\cal T}}\,e^{-Jt}
\sqrt{\frac{D}{\pi t}},
\end{eqnarray}
For $x>0$ and $0\leq \rho_0<1$ (with $t\to \infty$),  it is worth mentioning that there 
are, according to Eq. (\ref{CI}), two contributions to the long-time behavior of $c_{x}(t)-c_{x}(\infty)$: one, the subdominant, is given by Eq. (\ref{nx1d}) and the other, which always dominates reads \footnote{
Of course, in the case where $1-\rho_0 \ll 1$, this will require very long-time}:
\begin{eqnarray} 
\label{nx1CII}
 c_{x}(t)-c_{x}(\infty)&\simeq & -(1-\rho_0)\, e^{-Jt}
\left[{\rm erf} \left(
\frac{x}{2\sqrt{Dt}}
\right) + \frac{ e^{-x^2/4Dt}}{\frac{{\cal T}}{2}\sqrt{\frac{\pi t}{D}}}
\right],
\end{eqnarray}
where  ${\rm erf}(t)\equiv \frac{2}{\sqrt{\pi}} \int_{0}^{t} \, e^{-z^2} dz$
 is the usual error function \cite{Abramowitz}. Clearly, when  $0\leq \rho_0 <1$, the long-time behavior of the concentration could also have been obtained similarly as for Eq. (\ref{nx1d}).

From the results (\ref{nx1d}), (\ref{nx1CI}), (\ref{nx1CII}), it follows 
that in one-dimension the relaxation toward the stationary state is 
exponential, with a relaxation-time $J^{-1}$ and with a subdominant power-law prefactor 
$ \propto t^{-\nu(\rho_0)}$  which depends, as well as the amplitude, on the initial state of the system. In fact, $\nu(\rho_0=1)=\frac{3}{2}$ and 
$\nu(0\leq\rho_0<1)=\frac{1}{2}$. In this sense, the long-time dynamics is said to be non-universal.
 Other one-dimensional reaction-diffusion systems  displaying this kind of 
dynamical  behavior in the presence of a source (but without inhomogeneities) were studied \cite{DPAC}.

\section{Two-dimensional concentration in the presence of the source and the trap}
In this section we consider the two-dimensional situation in the thermodynamic limit ($L\to \infty$) and compute the
concentration profile in the presence of the source.
We first mention that in $2D$ and at $x=0$, for $J\geq 0$, one has 
${\cal I}_{\bm 0}(s)=\frac{2}{\pi} \, \frac{1}{(s+J+4D)}\, K\left(\frac{4D}{s+J+4D}\right)$,
where $K(y)\equiv \int_{0}^{\pi/2}
\frac{d\theta}{\sqrt{1-y^{2}\sin^{2}{\theta}}}$ is the  complete elliptic
  function of the first kind (see e.g. \cite{Itz}). 
It follows from the properties of this function that, as long as $J>0$,
${\cal I}_{\bm 0}(s)$ is well defined for all the values of $s$, including
$s\rightarrow 0$. From the previous remarks we infer  
 the stationary concentration at the origin of the lattice :
\begin{eqnarray}
\label{cstat0}
c_{0}(\infty)=\left[1+\frac{ {\cal T}}{\pi(J+4D)}\,
  K\left(\frac{4D}{J+4D}\right)\right]^{-1}.
\end{eqnarray}
Hereafter, we focus especially on the case where
the source and the trap are in {\it effective} competition, {\it i.e.} when 
the rate of the homogeneous source is ``small''
compared to the strength of the trap and to the diffusion rate.
In the limit of a weak source, {\it i.e.} when 
$J/D\ll 1$ and $J/{\cal T} \ll 1$ with $J>0$,  we have
$K\left([1+J/4D]^{-1}\right)\simeq \frac{1}{2}\ln{\left(D/J\right)}$
\cite{Abramowitz} and therefore $
 c_{0}(\infty)=
\frac{4\pi D}{{\cal T}\ln{(D/J)}}\left\{1+{\cal
    O}\left(1/\ln{[D/J]}\right)\right\}$.
This result shows that the concentration 
at the origin decays logarithmically with the strength of the source when the latter is small. 
We can also notice that for a perfect trap (${\cal T}/D\to \infty$ and ${\cal T}/J\to \infty$), the concentration at 
the origin vanishes: $c_{0}(\infty)=0$.
To compute the concentration of particles
at sites $x>0$, one has to evaluate ${\cal I}_{\bm x}(0)$.
Following the same reasoning as in the previous section, we notice
that, for $J>0$, the stationary concentration of particles
 is given by (\ref{cstat}), where one has to evaluate the 2D lattice-Green functions (\ref{Wat}) at $s=0$. 
In the physically interesting situation  when  the
competition between the weak source and the trapping reaction is really
effective, {\it i.e.} when $J/D \ll 1$
 and $ J/{\cal T} \ll 1$, with $J>0$, the main
contribution to ${\cal I}_{\bm x}(0)$ arises  from 
 the  $q\rightarrow 0$ expansion of the integrand of (\ref{Wat}), and one thus obtains (for $x$ large enough):
\begin{eqnarray}
\label{Ix2d}
{\cal I}_{\bm x}(0) \stackrel{J/D \ll 1}{\longrightarrow}\int_{-\pi}^{\pi }\int_{-\pi}^{\pi } 
\frac{dq_1 \, dq_2}{(2\pi)^2} \frac{e^{i {\bm q}. {\bm
      x}}}{J+Dq^{2}}
= \frac{1}{2\pi D}\, K_{0}\left(x\sqrt{J/D}\right),
\end{eqnarray}
where 
 $K_{0}(z)$ is the Bessel function of third kind \cite{Abramowitz} which behavior for small and large arguments
 is respectively
   $K_{0}(z) \stackrel{z\to 0}{\longrightarrow}
\ln{(1/z)}$ and $K_{0}(z)\stackrel{|z|\gg 1}{\longrightarrow}
(\pi/2z)^{1/2}\,e^{-z}$. Using these properties in the expression
(\ref{cstat}) together with the fact that ${\cal I}_{\bm 0}(0)\stackrel{J/D \ll 1 }{\longrightarrow} \ln{(D/J)}/4\pi D $, one obtains the two-dimensional stationary concentration
 in presence of  a {\it weak source}, which reads (for $x>0$ large enough):
\begin{eqnarray}
\label{concstat2D}
c_{x}(\infty)=1-\frac{2 K_0 \left(x\sqrt{J/D}\right)}{\ln(D/J)}. 
\end{eqnarray}
We notice that this expression is independent of the strength
${\cal T}$ of the trap. 
Again, to discuss the 
 perturbation due to the trap one has to distinguish two regimes ($x>0$):
 (i) when $0 \leq u\equiv x(J/D)^{1/2}\ll 1$, the concentration of particles increases logarithmically with the distance to the trap, with an amplitude proportional to $1/\ln{J^{-1}}$: $c_x(\infty)\simeq\ln{x^2}/\ln{(D/J)}  $; (ii) far away from the trap, when $u$ is large enough, the effect of the local perturbation vanishes exponentially with the distance to the origin: $\delta c_x(\infty)\simeq \sqrt{\pi/2u}\,e^{-u}/\ln{(D/J)}$. 
These results indicate that around the origin the competition between the (weak) source and the trap generates 
again a depletion zone (a circle) of radius $\ell_{2D}\sim (D/J)^{1/2}$. However the properties of the concentration profile are significantly different from those of the one-dimensional case. In fact, because of the subdominant $\sqrt{\pi/2u}$ prefactor, outside of the depletion region, when $x\gg \ell_{2D}$,  the effect of the perturbation decays faster with the distance to the trap than in 1D (where it decays merely as $e^{-u}$). The properties of the depletion region can be investigated further 
 by computing the previously introduced 
static ``$\theta_s$-distance'', $x_{\theta_{2D}}(\infty)$, which is 
the distance from the origin
 where the stationary concentration is the fraction $\theta_s$, 
from (\ref{concstat2D}) we obtain ($J\ll D$ and $J\ll {\cal T}$):
\begin{eqnarray}
\label{xthetas2D}
x_{\theta_{2D}}(\infty)\sim \left(\frac{D}{J}\right)^{\theta_s/2}.
\end{eqnarray}
This result shows that in 2D the $x_{\theta_{2D}}(\infty)$ distance is a nonlinear function of $\theta_s$, in contrast to what 
happens in 1D (\ref{xthetas1d}) and it is smaller than in the one-dimensional case: 
$x_{\theta_{1D}}(\infty)>x_{\theta_{2D}}(\infty)$.
From (\ref{cstat1d}) and (\ref{concstat2D}), we can also compare the effect of the trap  at the edge of the one and two-dimensional depletion zones in the limit of a weak source ({\it i.e.} for $x\sim (D/J)^{1/2}\gg 1$)
by computing $\frac{\delta c_{x}(\infty)|_{1D}}{\delta c_{x}(\infty)|_{2D}} \sim u^{1/2}\,\ln{J^{-1}} \gg 1$. This means that the difference of the effect of the perturbation due to the trap between the one and two-dimensional cases increases logarithmically with $J^{-1}$.

To compute the long-time behavior of the concentration within the depletion zone, we first consider
the case where initially $\rho_0=1$. In this situation, from Eq. (\ref{nx(s)}), we obtain
${\hat c}_{x}(s) \stackrel{s, \frac{J}{D} \ll 1}{\longrightarrow} -\frac{\ln{x^2}}{s\ln{(s+J)/D}}$.
After  Laplace-inversion, we obtain:
\begin{eqnarray}
\label{nx2d}
c_{x}(t)- c_{x}(\infty)\sim 
c_{ 0}(t)- c_{ 0}(\infty) \sim
\frac{e^{-Jt}}{Jt\, (\ln{Dt})^2}. 
\end{eqnarray}

For homogeneous, yet random, initial conditions with $0 \leq \rho_0<1$,
the main contribution to the long-time approach toward the stationarity,
within the depletion zone, is obtained from (\ref{n0(t)2dss}),
 (\ref{nx(t)2dss}) according  (\ref{CI}) :

\begin{eqnarray}
\label{nx2dCI}
c_{0}(t)- c_{0}(\infty)\simeq - (1-\rho_0) \,
\frac{4\pi D}{\cal T}
\frac{ e^{-Jt}}{\ln{(Dt)}}, 
\end{eqnarray}
and for $x\neq  0$ large enough, we have ($0 \leq \rho_0<1$):
\begin{eqnarray}
\label{nx2dCII}
c_{x}(t)- c_{x}(\infty)\simeq -(1-\rho_0) \,
 \frac{e^{-Jt} \ln{x^2}}{\ln{(Dt)}}. 
\end{eqnarray}
The long-time behaviors (\ref{nx2d}), (\ref{nx2dCI}) and (\ref{nx2dCII}) show
 that the approach toward
the stationary concentration is again exponential, with a relaxation 
time $J^{-1} \gg 1$. Since the dynamical prefactors  and the amplitude depend on $\rho_0$, the long-time dynamics 
is sensitive to the initial state of the system. In contrast to the one-dimensional case 
these subdominant prefactors are not power-laws but  
logarithmic terms, which means that in 2D the approach toward the steady-state is slower than in the one-dimensional case [see (\ref{nx1CI}) and (\ref{nx1CII})].

\section{Three-dimensional concentration in the presence of the source and the trap}
In the thermodynamic limit ($L\to \infty$), we now consider the three-dimensional situation.
In this case the integrals 
${\cal I}_{\bm x}(s)$ [see Eq.(\ref{Wat})] are always well defined, 
both when  $s\rightarrow 0$ and/or
$J\rightarrow 0$ ({\it i.e.} $\lim_{J\rightarrow 0}  {\cal I}_{\bm x}(0)<\infty$).
As a consequence, for all the values of $J\geq 0$ the concentration
profile evolves toward a reactive stationary state (with an infinite number
of particles). This is related to the fact that in dimensions
$d\geq 3$ random-walks are {\it transient} \cite{Privman,Itz} 
and therefore particles have a finite probability to be never trapped.
 
It turns out that the most fruitful approach to deal with the 3D case 
is to adopt a continuum formulation suitably combined with some exact discrete results.
In the continuum limit, one can indeed take advantage of the fact that
 the continuum differential equation 
 \begin{eqnarray}
 \label{eq}
 \left( \Delta -\kappa^{2} \right)N({\bm x})= F({\bm x}),
 \end{eqnarray}
where $\Delta$ is the continuum three-dimensional Laplacian and $F({\bm x})$ a known function, can be solved by the usual Green's function 
techniques :
one has (see, e.g., \cite{Carslaw,Book}) $N({\bm x})=-\frac{1}{4\pi} \int d^{3}{\bm x'} \, \frac{e^{-\kappa|{\bm
      x} - {\bm x'}|}}{|{\bm x} - {\bm x'}|} \, F({\bm x'})$.
As explained above, in the presence of a source $J\geq 0$, the  stationary concentration of particles is given by
Eq. (\ref{cstat}) and an explicit expression of the concentration (on the discrete lattice) requires the computation of the $3D$ integrals ${\cal I}_{\bm  x}(0)$, with $J\geq 0$ and $x>0$.  This is a hard task, and recently 
Delves and Joyce \cite{Delves}  have been able to calculate various  properties of the simple cubic
lattice  Green functions (\ref{Wat}), at ${\bm x} ={\bm  0}$ and for arbitrary values of
$J$. In particular, it follows from Eq.(6.12) of Reference \cite{Delves}:
\begin{eqnarray}
\label{I0}
{\cal I}_{\bm 0}(0)=
\frac{2(J+6D)-\sqrt{J(J+12D)}}{J^2 + 48D^2 +12JD} \;
 \left[_2F_{1}\left(\frac{1}{8}, \frac{3}{8}; 1; v\right)\right]^2,
\end{eqnarray}
where 
 $z\equiv \left(\frac{1}{3+\frac{J}{2D}}\right)^2$, $v\equiv \frac{16
     z[1-5z-(1-z)\sqrt{1-9z}]^2}{(1+3z)^4}$  and  
$ _2F_{1}\left(a , b ; c ; d \right)$ denotes Gauss hypergeometric
 function \cite{Abramowitz}. It follows from (\ref{n0(s)}),
 and (\ref{I0}) that the $3D$ stationary
 concentration of particles
 at the origin reads  (for $0\leq \rho_0
\leq 1 $):

\begin{eqnarray} 
\label{n03d}
c_{0}(\infty)=\frac{1}{1+{\cal T} {\cal I}_{\bm 0}(0)}.
\end{eqnarray}
Again, one can check that for a perfect trap (${\cal T}\to \infty$, with 
$D$ and $J$ finite), $c_{0}(\infty)=0$. 
We now turn to the study of the 3D stationary concentration for $x>0$ and $J>0$.
 To do this, we adopt the abovementioned reformulation which has already proven to be
 fruitful in some cases (see \cite{IVM,MI} and also Appendix A).
One therefore considers  $N({\bm x})\equiv c({\bm x}, \infty) -1$,
 where $c({\bm x},
\infty)$ is the solution of the stationary equations (\ref{nxdot})
in the continuum limit. In so doing, one sees that  $N({\bm x})$ obeys the equation (\ref{eq}), with
$\kappa\equiv \sqrt{J/D}$ and $F({\bm x})\equiv \frac{{\cal T}}{D}\,
\frac{1}{1+{\cal T} {\cal I}_{\bm 0}(0)}\, \delta^{3}({\bm x})$, where the three-dimensional Dirac delta
 function $\delta^{3}({\bm x})$ has been introduced.
From the solution of Eq.(\ref{eq}), one infers the following stationary
concentration of particles (for $x>0$) :
\begin{eqnarray}
\label{cstat3d}
c_{x}(\infty) \stackrel{x\gg 1}{\longrightarrow}
c({\bm x}, \infty)= 1-
 \frac{{\cal T}}{4\pi D\left[1+{\cal
    T}{\cal I}_{\bm 0}(0)\right]} \;\;\frac{e^{-x\sqrt{J/D} }}{x}. \nonumber\\
\end{eqnarray}

Eq. (\ref{cstat3d}), together with  (\ref{I0}) and (\ref{n03d}), gives an explicit
 expression  of the  $3D$ concentration of particles.
In fact, the result (\ref{cstat3d}) is  the exact continuum $3D$ stationary
 concentration of particles, valid for any values of $J>0$ and is in excellent 
 agreement with the stationary  discrete solution of (\ref{nxdot}).
 
The result (\ref{cstat3d}) shows that in 3D the effect of the trap 
is much ``softer'' than in low dimensions: as the spatial perturbation of the static concentration 
due to the trap is $\propto  e^{-x\sqrt{J/D}}/x$,
there is only a local deviation of the concentration from the value one rather than a real depletion, contrary to what happens in low  dimensions. In fact, the properties of the 3D concentration profile significantly differ from those 
obtained in 1D and 2D: (i) for a weak source, {\it i.e.} 
$J/D\ll 1$ and $J/{\cal T}\ll 1$, the concentration approaches the value 1 as  $\delta c_{x}(\infty)\propto x^{-1}$;  (ii) while, when $u\gg 1$ the effect of the perturbation is not merely exponential with the distance to the trap: $\delta c_{x}(\infty)\propto e^{-u }/x$.
 To compare the effect of the trap in 3D with the one and two-dimensional cases in the limit of a weak source, 
  using (\ref{cstat1d}), (\ref{concstat2D}) and (\ref{cstat3d}), one can compute the ratio $r_{13}={\rm lim}_{x\gg 1, J/D \ll 1, J/{\cal T}\ll 1}\frac{\delta c_{x}(\infty)|_{1D}}{\delta c_{x}(\infty)|_{3D}}$ between $\delta c_{x}(\infty)$ in 1D and 3D and  
  $r_{23}={\rm lim}_{x\gg 1, J/D \ll 1, J/{\cal T}\ll 1}\frac{\delta c_{x}(\infty)|_{2D}}{\delta c_{x}(\infty)|_{3D}}$ between the two and three-dimensional cases, and one finds:
 $r_{13}\sim r_{23}\propto x\gg 1$. These results confirm that the effect of the trap on the stationary concentration profile is 
 much more important in low dimensions, where the random walks are recurrent, than in 3D where they are transient.

As previously, to determine rigorously the approach toward the stationary
concentration (\ref{cstat3d}) one should analyze the $s\to 0$ behavior of 
(\ref{nx(s)}), which is difficult. However, it follows from Eq.(\ref{sol}) that for an initially completely occupied lattice ($\rho_0=1$)
we have: $c_{x}(t)-c_{x}(\infty) \sim e^{-Jt}\,
t^{-3/2}$. For $0\leq \rho_0<1$, it still follows from  (\ref{nx(s)}) and
from (\ref{CI}), that $c_{x}(t)-c_{x}(\infty) \sim e^{-Jt}\,
t^{-1/2}$. We therefore concludes that also in 3D the perturbed long-time dynamics is sensitive to the initial state of the system.

\section{Global effect of  the trap and the source}

In the previous sections the  effects of trapping
and injection reactions have been studied via the  computation of concentration
 at each site of the lattice. 
Another question which is also relevant is related to the 
global effect of the source and the trap. As in the absence of the trap the system is fully occupied at $t\to \infty$, it is interesting to compute the total number  $B(\infty)$ of particles having been removed from the lattice by the trap and not having been replaced by the source. 
 To address this issue, and thus to gain some insight on the global effect of 
the competition between the trapping and injection reactions, one computes: $B(t)\equiv
\sum_{x_1=-\infty}^{\infty}\dots \sum_{x_d=-\infty}^{\infty}
(1- c_{ x}(t))$, which provides the number of particles that have been removed from the system (without having been replaced) up to a given time $t$.
It follows from  Eq. (\ref{sol}): 
\begin{eqnarray} 
\label{B(t)}
 B(t)=
 {\cal T} e^{-Jt}\int_{0}^{t} d\tau \, e^{J \tau} c_{ 0}(\tau).
\end{eqnarray}

Without help of any specific knowledge of $c_{0}(t)$,  as  one
  always has $0 \leq c_{x}(t)\leq 1$, we readily obtain the
  following bounds: $0\leq B(t)\leq \frac{{\cal T}}{J}(1-e^{-Jt})$, and thus $0 \leq  B(\infty)\leq 
  \frac{{\cal T}}{J}$.
It therefore follows from (\ref{B(t)}), that in the long-time
regime:
\begin{eqnarray}
\label{Blong12d}
B(t)= \frac{{\cal T}}{J} \, c_{ 0}(\infty) [1-e^{-Jt}],   
\end{eqnarray}
where $c_{0}(\infty)$ has been computed in the previous sections. This shows that $B(t)$ reaches the
 stationary value $B(\infty)=\frac{{\cal T}}{J}c_{\bm 0}(\infty)$
  exponentially fast, with a relaxation-time $J^{-1}$.
Let us now focus on the study of  $B(\infty)$. 
 With (\ref{cstat1d}), (\ref{cstat0}), (\ref{n03d}) and (\ref{I0}) we obtain:
 
\begin{eqnarray}
\label{Binfty}
 B(\infty)= \cases{ 
\frac{{\cal T}}{\sqrt{J}} \, \frac{\sqrt{J+4D}}{{\cal T} + \sqrt{J(J+4D)}}
&; in 1D \\  
\frac{{\cal T}}{J} \, \left[1+\frac{{\cal T}  }{\pi (J+2D)} K\left(
\frac{4D}{J+4D}
\right)\right]^{-1}  &; in 2D \\
\frac{{\cal T}}{J[1+{\cal T} {\cal I}_{\bm 0}(0)]} &; in 3D.}
\end{eqnarray}
It is therefore clear from (\ref{Binfty}) that the competition between the
source and the trap  also depends  {\it
  globally}, in a non-trivial manner, on the spatial dimension of the system.
In particular, in the limit of a  weak source ($J/D\ll 1, J/{\cal T}\ll 1$),
 $B(\infty)$  diverges as $J^{-1/2}$ in 1D, as $(J\ln{J})^{-1}$ in 2D and as $J^{-1}$ in 3D.
As a consequence, we can compute the way in which the intensity of the source should vanish to allow the trap to remove a number of particles comparable to the size of the system ($L^{d}$, $L\to \infty$): it turns out that in 1D 
one should have $J/D\sim L^{-2}$, in 2D $J/D\sim (L^{2}\ln{L})^{-1}$ and $J/{\cal T}\sim L^{-3}$.

\section{Discussion and Conclusion}

In this work, one has studied, in dimensions $d=1, 2$ and $3$, mainly in the thermodynamic limit, 
 the kinetics of a simple stochastic many-body
system where particles, subject to volume exclusion, diffuse and interact with a
 localized trap in the  presence of an external uniform source refilling the empty sites of the lattice. It has been shown that the competition between the homogeneous and local 
reactions always gives rise to  non-trivial reactive (fluctuating) and genuine nonequilibrium steady states. 

In fact, in 1D, an explicit expression for the stationary concentration profile of particles on an infinite chain has been obtained and it has been shown that the density of particles deviates from the value one exponentially with the distance to the trap (as $\propto e^{-\xi x}$), while in 2D, in the thermodynamic limit, the concentration has been computed in the limit of a weak source (compared to the other processes), where the interplay between the source and the trap is particularly effective. 
Interestingly, in 1D the steady-state displays the same qualitative behavior as the one obtained in the
one-dimensional $AA \leftrightarrow A$ reaction-diffusion system in the presence of a perfect trap \cite{benA}.
From these results it was inferred that in low dimensions the competition between the trap and the source 
generates a depletion zone in the concentration profile. Both in 1D and 2D these regions are isotropic around the trap and their ``radii'' are proportional to the inverse of the square root of the injection rate (when the latter is small). We have also quantitatively analyzed the differences in these static profiles: while in 1D the concentration within the depletion zone increases linearly, in 2D the spatial perturbation is less important and indeed
depends logarithmically on the distance to the trap.
In three dimensions, in the thermodynamic limit, it has been shown, using a suitable continuum limit reformulation, that the effect of 
the perturbation due to the trap results only in a deviation from the value one by a 
term $\propto e^{-x\sqrt{J/D} }/x$. In this case there is no depletion in the concentration profile and the trapping reaction is much less effective than in low dimensions: it just generates a short-range deviation from the value one.
To explain these different behaviors, one has to keep in mind the fact that random walks are recurrent in 
one and two dimensions, while they are transient in three dimensions (and above). Hence, in low dimensions, 
the particles on the lattice are doomed to visit the origin and thus might be trapped. This is no longer the case in 3D, where the particles have a finite probability to never visit the trapping site.  
The long-time approach of the concentration toward the
reacting steady state has also been computed and it was shown that the inhomogeneous dynamics is non universal and no 
longer purely exponential: there are nontrivial subdominant prefactors which depend on the initial state of the system. The latter are logarithmic functions of the time in 2D  and
  power-laws in 1D and 3D.

 In summary, the results of this work show that already a single
inhomogeneity may deeply, and non-trivially, affect in {\it all dimensions} the spatiotemporal properties of a  diffusive stochastic many-body system characterized by the competition between an uniform source and a local trap. As the 1D version of the system under consideration displays the same kind of behavior as another reaction-diffusion model perturbed by a trap, it might be that the results obtained for this simple system could also apply more generally to a class of systems where uniform and local reactions compete. In particular, it would be interesting to determine if models like the reaction-diffusion process $AA \leftrightarrow A$ in the presence of a trap would behave in the same manner as the model studied here also in two and three dimensions (and not only in 1D).

\ack
P. L. Krapivsky and S. Redner are thanked for many
valuable discussions. The author is also grateful to I. T. Georgiev, B. Schmittmann, 
U. C. T\"auber and R. K. P. Zia.
The financial support of the Swiss National Foundation through the
fellowship N. 81EL-68473 is acknowledged. This
work was also partially supported by US NSF DMR-0414122 and 0308548.

\appendix

\section{Diffusive stochastic system in the presence of a trap and in the absence of the source}

In the absence of the source, as the hopping process is symmetric and as there is no difference
in between exchanging two particles or leaving them in place, the site restriction plays no role on the physical properties of the concentration of particles.
Therefore, when $J=0$, the concentration of particles in the system can be studied via a model of noninteracting particles (simple random walkers)
obeying a diffusive equation supplemented by a radiative term \cite{Carslaw}.
Such a study has been carried out in the references \cite{Ben,Weiss,Taitelbaum2,Taitelbaum}.
Hereafter, for the sake of completeness one briefly outlines how these results can be derived on a discrete lattice (in the thermodynamic limit) by using the techniques described in this work. In  addition, as original results, the time necessary for 
trapping a macroscopic number of particles ({\it i.e} comparable to the system size) is also computed.

\subsection{The one-dimensional case}
Here, in the limit $L\to \infty$, we consider the one-dimensional situation without the source 
($J=0$) and in this case the system reduces to a so-called ``symmetric exclusion process'' (see \cite{Spitzer,Schutzrev}) perturbed by a trap. Because of the sole presence of the trap, and as particles
perform basically random walks,
the  concentration is clearly doomed to
vanish at long-time ($t\rightarrow \infty$).  
Here, one outlines the steps allowing to recover, starting from a discrete formulation and using other methods, the results already reported in Refs. \cite{Ben,Weiss,Taitelbaum2,Taitelbaum}. 
To address this point one analyzes Eq. (\ref{nx(s)}) in the
long-time regime, where $s\rightarrow 0$.
It follows from (\ref{Ix1d}) that $ {\cal I}_{0}(s)\stackrel{s\rightarrow 0}{\longrightarrow} \frac{1}{2\sqrt{Ds}}$,
and, for $s\rightarrow 0$  with $x\sqrt{s}$ kept fixed, we have 
$ {\cal I}_{x}(s)\stackrel{s\rightarrow 0}{\longrightarrow}  \frac{e^{-x\sqrt{s/D}}}{2\sqrt{Ds}}$.
Inserting these expressions into the general formula (\ref{nx(s)})
one obtains $\hat{c}_{0}(s)\stackrel{s\rightarrow 0}{\longrightarrow}  \frac{2}{{\cal T}\sqrt{s/D}}$
and  $\hat{c}_{x}(s)\stackrel{s\rightarrow 0}{\longrightarrow}  \frac{\rho_0}{s}
\left(
1-\frac{e^{-x\sqrt{s/D}}}{1+\frac{2\sqrt{D s}}{{\cal T}}}
\right)$. Therefore, after Laplace-inversion \cite{Abramowitz}, one gets
the following long-time behavior of the concentration 
($Dt\rightarrow \infty$ and ${\cal T}t\rightarrow
 \infty $) :
$  c_{0}(t)= \frac{2\rho_0}{{\cal T}} \, \sqrt{\frac{D}{\pi t}}$, and more generally:
\begin{eqnarray} 
\label{nx(t)1dss}
  c_{x}(t)&=&  \rho_0 \;{\rm erf} \left(
\frac{x}{2\sqrt{Dt}}
\right) + \rho_0 \;e^{\frac{{\cal T}}{2D} \left(
x+{\cal T}t/2
\right)  }
 {\rm erfc} \left(
\frac{x}{2\sqrt{Dt}}
+ \frac{{\cal T}}{2}\sqrt{\frac{t}{D}}
\right)\nonumber\\
&\simeq&
\rho_0 \;\left[{\rm erf} \left(
\frac{x}{2\sqrt{Dt}}
\right) + \frac{ e^{-x^2/4Dt}}{\frac{{\cal T}}{2}\sqrt{\frac{\pi t}{D}}}
\right], 
\end{eqnarray}
where ${\rm erfc}(t)\equiv 1-
{\rm erf}(t)$ is the complementary error function \cite{Abramowitz}.
It is worth to notice that this result coincides with Eq. (4)
 of Reference 
\cite{Phototrap} (see also \cite{Ben,Weiss,Taitelbaum2}).
The expression (\ref{nx(t)1dss}) is particularly appealing in the scaling
 regime where $x\rightarrow \infty$  and  $t\rightarrow \infty$, with
$x^{2}/t$ kept finite. In fact it is readily checked that  for
 $x/\sqrt{Dt} \rightarrow 0$, one has $c_{x}(t)\rightarrow 0$, 
whereas for large time and at sites very distant from the trap, i.e. $x\gg
 \sqrt{Dt}\gg 1$,  $c_{x}(t)\rightarrow \rho_0$. This means that 
there is an isotropic depletion zone around the trap. The symmetric 
fronts of this region propagate as $t^{1/2}$. Inside the depletion area, the
concentration vanishes algebraically with the time and from (\ref{nx(t)1dss}) one gets :
\begin{eqnarray} 
\label{nx(t)1dssdepl}
  c_{x}(t)=  
\frac{\rho_0 \, \{x + \frac{2D}{{\cal T}}\}}{\sqrt{\pi \,Dt}}.
\end{eqnarray}
This result shows that within the depletion zone 
the concentration of particles vanishes as $t^{-1/2}$. 
Let us also recall that a quantity which is theoretically relevant \cite{Ben,Taitelbaum2} and, 
more importantly, which has been
 measured experimentally \cite{Phototrap} is the so-called 
``$\theta$-distance'' ($x_{\theta}$). This is the distance from the trap to the point 
where the concentration of $A$ particles reaches the specific fraction
 $\theta$  ($0 \leq \theta\leq 1$) of its value in the bulk \cite{Phototrap}.
Requiring that  $c_{x}(t)= \theta\, \rho_0$ on the l.h.s of 
 Eq. (\ref{nx(t)1dssdepl}), one obtains:
\begin{eqnarray} 
\label{xthetass1d}
 x_{\theta}(t)= \theta \sqrt{\pi \,Dt}-\frac{2D}{{\cal T}}.
\end{eqnarray}

 In recent experiments (and numerical simulations), the depletion zone 
was indeed found to grow as the square root of the time 
\cite{Phototrap};
 in agreement  with the theoretical
  predictions (\ref{nx(t)1dss}),
 (\ref{nx(t)1dssdepl}) and (\ref{xthetass1d})
\cite{Phototrap,Ben}.
 From Eq.(\ref{sol}), we can also notice  
that the total number of particles removed from the system by the trap grows
as $t^{1/2}$ in 1D, which means that the time $T$ necessary to eliminate a number of particles
comparable to the system size is $T\sim L^2$, with $L\to \infty$.

\subsection{The two-dimensional case}
We now turn to the two-dimensional situation in the absence of the
source, i.e. $J=0$ and in the limit where $L\to \infty$. As there is no external injection of particles and random walks are
 known to be still recurrent in 2D the final concentration of particles vanishes, like in 1D.
As in the one-dimensional case, here the issue is mainly technical as it aims at showing
 how to reproduce from a discrete formulation the spatiotemporal structure of the
concentration profile, already obtained in Refs \cite{Ben,Weiss,Taitelbaum2,Taitelbaum}.

Using the fact that the complete elliptic functions and  ${\cal I}_{x}(s)$ diverge 
logarithmically as
$K\left(\frac{1}{1+s/4D}\right)\stackrel{s\rightarrow 0}{\longrightarrow}-\ln{s}$;
${\cal I}_{0}(s)\stackrel{s\rightarrow 0}{\longrightarrow} -\frac{1}{4\pi D} \,
 \ln{\left(s/D\right)},$
\cite{Abramowitz}, and ($x\gg 1$)
${\cal I}_{\bm x}(s)\stackrel{s\rightarrow 0}{\longrightarrow}
  \frac{1}{2\pi D}\, K_{0}\left(x\sqrt{s/D}\right)$, 
we obtain:
  $\hat{c}_{0}(s)\stackrel{s\rightarrow 0}{\longrightarrow} -\frac{4\pi D}{{\cal T}}
\frac{\rho_0}{s \,\ln{(s/D)}}$ and
 $ \hat{c}_{x}(s)\stackrel{x^2 s, \frac{s}{{\cal T}}\rightarrow 0}{\longrightarrow}-\rho_0 \, \frac{\ln{x^2}}{s \, \ln{(s/D)}}$.
Therefore, after Laplace-inverting these expressions, one gets 
the following long-time behavior of the concentration ($Dt\rightarrow \infty$):
\begin{eqnarray} 
\label{n0(t)2dss}
  c_{0}(t)=  \frac{4\pi \rho_0 D}{{\cal T}} \, \frac{1}{\ln{(Dt)}},
\end{eqnarray}
and,  $Dt, {\cal T}t \gg x^2 \gg 1$,  
\begin{eqnarray} 
\label{nx(t)2dss}
  c_{x}(t)= \rho_0 \, \frac{\ln{x^2}}{ \ln{(Dt)}}.
\end{eqnarray}
These results show that in two dimensions, and in the absence of the source, 
 the concentration at sites $x\ll t^{1/2}$ still vanishes  
at long-time. More precisely, 
there is an isotropic depletion zone of time-dependent 
radius $\ell(t)\sim t^{1/2}$ around the origin.
Within this region, the concentration of particles decays logarithmically with the time
 [$\propto 1/\ln(t)$], {\it i.e.} much slower than in 1D
 [see (\ref{nx(t)1dssdepl}) and (\ref{nx(t)2dss})].
As a consequence, at variance with  the 1D situation, because of the
 logarithmic long-time dependence of the concentration, the
two-dimensional dynamics 
{\it does not exhibit   scaling}. This point is illustrated further
 by the computation of the 
previously introduced ``$\theta$-distance'',
 $x_{\theta}(t)$, which has also very recently been
 measured experimentally \cite{Photobleach2D}. In fact,
 plugging $c(t)=\theta \, \rho_{0}$ on the l.h.s of (\ref{nx(t)2dss}), one
 obtains ($Dt\gg 1$, ${\cal T}t \gg 1$):
\begin{eqnarray} 
\label{xtheta2d}
  x_{\theta}(t)\sim  (Dt)^{\alpha_{2}}, \; \alpha_{2}=\theta/2.
\end{eqnarray}
This result, which is in agreement with the theoretical results obtained for a
continuum model \cite{Taitelbaum} and with
recent experimental and numerical results \cite{Photobleach2D}, shows 
that the  $\theta$-distance at which the concentration of particle
 is $\theta \,\rho_{0}$ evolves as a 
power-law characterized by 
the exponent $\alpha_{2}=\theta/2$. This result should be contrasted with
the result (\ref{xthetass1d}), where the exponent is simply $1/2$, independently of the value of $\theta$. Also, it follows from (\ref{sol}) 
that the total number of particles removed from the system by the trap grows
as $t/\ln{t}$ in 2D, which implies that the time $T$ necessary to trap and remove a number of particles
proportional to system size is $T\sim L^{2}\ln{L}$, with $L\to \infty$.
 
 \subsection{The three-dimensional case}
The three-dimensional case, which has been less studied \cite{Taitelbaum2}, displays results which are qualitatively different from the low-dimensional cases already analyzed. Here we consider the thermodynamic limit $L\to \infty$.

As discussed above, in three dimensions, even in the absence of the source, i.e. when $J=0$,
the quantity  ${\cal I}_{{\bm x}}(s=0)$ is well defined. 
It follows from Eqs. (\ref{nx(s)}) and (\ref{CI}) that the stationary
concentration of particles in the absence of source reads:
\begin{eqnarray}
\label{cstat3dss}
c_{x}(\infty)=\rho_0 \, \left[
1-\frac{{\cal T}{\cal I}_{\bm x}(0) }{
1+ {\cal T}{\cal I}_{\bm 0}(0)
}\right].
\end{eqnarray}
To make this expression explicit, we take advantage of the recent results obtained by Glasser and
Boersma \cite{Glasser}, who have been able to give the exact expression of 
 ${\cal I}_{\bm x}(0)$. Following their work, we introduce the quantity
$g_0\equiv \frac{\sqrt{3}-1}{96 \pi^3} \, \Gamma^2\left(\frac{1}{24}\right)
\Gamma^2\left(\frac{11}{24}\right)= 0.505462...$, where $\Gamma(z)$ is the
usual Euler's Gamma function \cite{Abramowitz}. According to Reference
\cite{Glasser}, we also introduce a triplet $(u_{{\bm x}}, v_{{\bm x}},
w_{{\bm x}})$ of rational numbers, depending on ${\bm x}$, and whose value is
given [for ${\bm x}=(1, 0, 0)$ to ${\bm x}=(5, 5, 5)$] in the Table 2 of
 Reference \cite{Glasser}. 
With these notations, and with help of the results obtained in
\cite{Glasser}, the expression (\ref{cstat3dss}) explicitly reads:
\begin{eqnarray}
\label{cstat3dssexpl}
c_{ x}(\infty)=\rho_0 \, \left[
1-
\frac{{\cal T}\left( u_{{\bm x}} g_0^2 +  w_{{\bm x}} g_0 + \frac{v_{{\bm
          x}}}{\pi^2}\right)}{g_0 \,(2D + {\cal T}g_0)}
\right].
\end{eqnarray}
We remark that for a perfect trap, i.e. when ${\cal T}\rightarrow
\infty$, (\ref{cstat3dssexpl}) reduces to an expression that is independent of
$D$:
$c_{ x}(\infty)=\rho_0 \, \left[
1-
\frac{ u_{{\bm  x}} g_0^2 +  w_{{\bm x}} g_0 + \frac{v_{{\bm
          x}}}{\pi^2}}{g_0^2}
\right]$.
It is useful for the sequel to have explicitly the expression of the
concentration at the origin, where $(u_{{\bm 0}}, v_{{\bm 0}}, w_{{\bm
    0}})= (1, 0, 0)$ \cite{Glasser}. With (\ref{cstat3dssexpl}), we obtain:
\begin{eqnarray}
\label{cstat3d0}
c_{ 0}(\infty)=
\frac{2D \, \rho_0}{2D + {\cal T}g_0 }.
\end{eqnarray}
Clearly, for a perfect trap (${\cal T}\to \infty$ and $D$ finite),
 one has $c_{ 0}(\infty)=0$. As  only a finite number of triplets  $(u_{{\bm x}}, v_{{\bm x}},
w_{{\bm x}})$ are  given in \cite{Glasser}, and as their expression and
computation are non-trivial, to gain some insight on the 
stationary concentration in a simple manner, it
 is fruitful to adopt 
a continuum reformulation, as already devised in References \cite{IVM,MI}. 
One therefore considers the quantity $N({\bm x})\equiv c({\bm x}; \infty)
-\rho_0 $, where $c({\bm x}; \infty)$ is the stationary concentration in the
continuum limit. According to the continuum limit of the stationary version of
Eq (\ref{nxdot}), $N({\bm x})$ obeys the differential
equation (\ref{eq}), with $\kappa =0$ and $F({\bm x})= \frac{{\cal T}}{D}
c_{0}(\infty) \delta^{3}({\bm x})$.
In this case, the equation (\ref{eq}) for $N({\bm x})$ is reduced to the
problem of determining the ``electrostatic potential'' due to the ``charge''
 ${\cal T}c_{ 0}(\infty)/D$ located at the origin. The ``charge'' at the origin is 
 therefore got from $c_{ 0}(\infty)$ (\ref{cstat3d0}) and one obtains the following
expression of the stationary concentration in the continuum limit (see \cite{IVM,MI}):
\begin{eqnarray}
\label{cstat3dcont}
c( {\bm x},\infty)= \rho_0 \, \left[
1-\frac{{\cal T}}{2\pi (2D+ {\cal T}g_0)} \, \frac{1}{x} \right],
\;\; (x>0).
\end{eqnarray}
It has been checked that this continuum expression is in excellent agreement with the
 exact discrete result (\ref{cstat3dssexpl}) and even ``coincides'' with the
latter for $x\gg 1$.  
It  follows from this discussion that in 3D,
for $x>0$ and $J=0$, the stationary concentration is a 
smooth and isotropic function
that evolves (for $x>0$)
 as the inverse of the distance to the origin [the value of the
 concentration at the origin is given by (\ref{cstat3d0})]: 
$c_{{x}}(\infty) \approx c({\bm x}, \infty)=\rho_0[1- {\cal A}(D, {\cal T})/x]$, where 
the amplitude ${\cal A}(D, {\cal T})\equiv\frac{ {\cal T}}{2\pi (2D+ {\cal T}g_0)} $
 depends explicitly on the strength of the trap. For a ``perfect'' trap, we
 have the same behavior but with an amplitude that reads 
$A(D, {\cal T}\rightarrow
 \infty)= 1/(2\pi\, g_0)$. At sites very far away from the trap, i.e $x\to \infty$, 
the stationary concentration is the initial one: $c_{{ x}}(\infty)
 \approx c({\bm x}, \infty)   \stackrel{x\rightarrow \infty}{\longrightarrow}   \rho_0$.
Also, from Eq. (\ref{sol}) one can show that the total number of particles removed from the system by the trap grows
$\propto t$ in 3D.
\section{Concentration in the presence of the source and the trap on periodic finite size lattices}
So far, this work has focused on infinite systems. However, it is well known that boundary conditions 
may play a crucial role in nonequilibrium statistical
systems. For instance, even for simple one-dimensional
 models such as the asymmetric exclusion process (ASEP) [see e.g. 
\cite{Privman} and references therein] open boundaries induce {\it phase transitions}. 

As the influence of boundary conditions is worth being dealt with care, in this appendix one briefly reformulates the problem under consideration for the case of {\it periodic boundary conditions} on  finite size lattices.

To emphasize the periodicity of the system, one slightly changes the notations and consider that the 
$N\equiv 2L+1$ sites  are labelled by a vector ${\bm x}$ of components $0 \leq x_i\leq N-1$.
Again, one considers that a trap of strength ${\cal T}$ is located at the origin of the lattice (site ${\bm 0}$). 

As it was mentioned, the equation of motion for the concentration of particles on periodic lattices is still given by Eq.(\ref{nxdot}).
 In the absence of the trap ({\it i.e.} ${\cal T}=0$), the solution of Eq.(\ref{nxdot}) can be obtained following 
 Glauber \cite{Glauber} and thus reads: $c_{\bm x}(t)|_{{\cal T}=0}=1+\sum_{{\bm y}}(c_{\bm y}(0)-1)e^{-Jt}\prod_{i=1}^{d}\left[e^{-2Dt} \sum_{n_i=-\infty}^{\infty} I_{x_i-y_i+n_iN}(2Dt)\right]$, where the summation
 over $n_i=-\infty, \dots, +\infty, \, i=1, \dots, d$ accounts for the $N-$periodicity in all the directions $i$.
 Following the same steps as in Section 3, one can therefore derive the following integral relationship obeyed by the concentration on periodic lattices of size $N$:

\begin{eqnarray}
\label{solper}
 c_{\bm x}(t) &=& 1+ \sum_{\bm y} (c_{\bm y}(0)-1) e^{-Jt}\prod_{i=1}^{d}
 \left[e^{-2Dt} \sum_{n_i=-\infty}^{\infty} I_{x_i-y_i+n_iN}(2Dt)\right] \nonumber\\
 &-&{\cal T} \int_{0}^{t} dt' c_{\bm 0}(t-t') \, e^{-Jt'} \prod_{i=1}^{d}
 \left[e^{-2Dt'} \sum_{n_i=-\infty}^{\infty}I_{x_i+n_iN}(2Dt')\right], \nonumber\\
\end{eqnarray}
Again, using the convolution theorem and the properties of Laplace transform, one obtains the 
expression for the Laplace transform of the concentration of particles and then proceeds as in Sections 4, 5 and 6 to obtain its explicit spatiotemporal expression in 1D, 2D and 3D, respectively.   

\subsection{Concentration of particles in the presence of the source and the trap on a finite ring}

For the sake of concreteness and simplicity, here one illustrates how to compute the 
concentration of particles on a ring of (finite) size $N=2L+1$, with an initial homogeneous concentration 
$c_{x}(0)=\rho_0$.

Proceeding as in Section 3, it follows directly from Eq.(\ref{solper}) that the Laplace transform of the concentration of particles reads  ($0 \leq x\leq N-1$)
\begin{eqnarray} 
\label{nx(s)per}
 \hat{c}_{x}(s)=
  \frac{1+{\cal T} \sum_{n=-\infty}^{\infty}[ {\cal I}_{nN}(s) -{\cal I}_{ x+nN}(s) ]}{1+{\cal T}
  \sum_{n=-\infty}^{\infty}{\cal I}_{nN}(s)}\left[\frac{1}{s} -\frac{1-\rho_0}{s+J}\right].
\end{eqnarray}
As a consequence, the stationary concentration  is given by ($0\leq x\leq  N-1$)
\begin{eqnarray} 
\label{cstatper}
 c_{x}(\infty)=
  1-\frac{{\cal T} \sum_{n=-\infty}^{\infty} {\cal I}_{x+nN}(0)}{1+{\cal T} \sum_{n=-\infty}^{\infty} {\cal I}_{nN}(0)}
\end{eqnarray}
To make use of Eq.(\ref{Ix1d}), one writes
\begin{eqnarray} 
\label{Id}
  \sum_{n=-\infty}^{\infty} {\cal I}_{x+nN}(s)=\sum_{n=0}^{\infty}\left[{\cal I}_{x+nN}(s)+{\cal I}_{(n+1)N-x}(s)\right].
\end{eqnarray}
Using (\ref{Id}) and Eq.(\ref{Ix1d}), recalling that $\xi\equiv 2\ln{\left(\frac{\sqrt{J}+\sqrt{J+4D}}{2\sqrt{D}}\right)}$, one obtains $\sum_{n=-\infty}^{\infty} 
{\cal I}_{x+nN}(0)=\frac{1}{\sqrt{J(J+4D)}}\, \left(\frac{e^{-\xi x}+ e^{-\xi(N-x)}}{1-e^{\xi N}}\right) $. Plugging this result into Eq. (\ref{cstatper}), one obtains the stationary concentration of particles on a periodic lattice of finite size $N$:
\begin{eqnarray} 
\label{ConsStatPer}
 c_{x}(\infty)=
  1-\frac{{\cal T} \left(e^{-\xi x}+ e^{-\xi(N-x)}  \right)}{{\cal T}+\sqrt{J(J+4D)}+
  \left[{\cal T}-\sqrt{J(J+4D)}\right]\,e^{-\xi N}}.
\end{eqnarray}
Plugging $x=0$ and $x=N$ into Eq. (\ref{ConsStatPer}) one obtains the same expression , as required by the periodicity (the sites $0$ and $N=2L+1$ now coincide). Clearly, we see that for a finite ring, there are differences with respect to the infinite chain, discussed in Section 4. For instance, when the strength of the source is weak ({\it i.e.} $J/{\cal T}\ll 1$ and $J/D\ll 1$) and $0 \leq x(J/D)^{1/2}\ll 1$, the stationary concentration now varies as $c_x(\infty)=x\, \xi \tanh{(\xi N/2)}$, {\it i.e.} the slope of the stationary concentration profile in the vicinity of the trap increases with the size $N$ of the lattice.
 In the thermodynamic limit, where $N\to \infty$, it is easy to check that the expression (\ref{cstatper}) simplifies and, as expected, one recovers the expression (\ref{cstat1d}) obtained for an infinite chain.
 
 To determine the long-time behavior of the concentration on a finite ring, one
  first makes use of Eqs.(\ref{Id}) and (\ref{Ix1d}), and thus finds (for $J=0$):
\begin{eqnarray} 
\label{Id2}
&& \frac{{\cal T} \sum_{n=-\infty}^{\infty}{\cal I}_{ x+nN}(s)|_{J=0}}{1+{\cal T}
  \sum_{n=-\infty}^{\infty}{\cal I}_{nN}(s)|_{J=0}} \nonumber\\
  &\stackrel{s\rightarrow 0}{\longrightarrow}&
   \frac{1}{1+\frac{2\sqrt{Ds}}{{\cal T}}} \sum_{k=0}^{\infty}(-1)^{k}\,\left[e^{-(x+kN)\sqrt{s/D}}+e^{-(N\{k+1\}-x)\sqrt{s/D}}\right]
\end{eqnarray}
 Following the same reasoning as in Section 4, it follows from Eq. (\ref{nx(s)per}) [see also Eq.(\ref{CI})] that for 
 an initially completely occupied system ($\rho_0=1$) the concentration of particles reads $c_{x}(t) =1- \int_{0}^{t} dt' e^{-Jt'} 
{\cal L}^{-1} \left[
\frac{{\cal T} \sum_{n=-\infty}^{\infty}{\cal I}_{ x+nN}(s)|_{J=0}}{1+{\cal T}
  \sum_{n=-\infty}^{\infty}{\cal I}_{nN}(s)|_{J=0}}
\right](t')$. Using this relationship together with Eq.(\ref{Id2}), one finds the long-time behavior ($0<x<\infty$ and $t\to \infty$):
\begin{eqnarray} 
\label{clongper1}
&&c_x(t)-c_x(\infty)\simeq \frac{e^{-Jt}}{2Jt\sqrt{\pi Dt}}\, \sum_{k=0}^{\infty}(-1)^{k} \nonumber\\
&\times&\left[
(x+kN)e^{-(x+kN)^{2}/4Dt}+\left\{N(k+1)-x\right\} e^{[N\{k+1\}-x]^2/4Dt}
\right]
\end{eqnarray}
 Following the same steps as in Section 4 and with help of (\ref{Id2}), one can also obtain the long-time behavior  ($t\to \infty$) of the concentration when $0\leq \rho_0<1$. According to Eq.(\ref{nx(s)per}) that there are two contributions to 
 $c_x(t)-c_x(\infty)$:  the subdominant one is given by the r.h.s. of Eq.(\ref{clongper1}), while the dominant contribution reads
\begin{eqnarray} 
\label{clongper2}
&&c_x(t)-c_x(\infty)\simeq -(1-\rho_0)e^{-Jt}\nonumber\\ &\times&\left[1- \sum_{k=0}^{\infty}(-1)^{k} \left\{
{\rm erfc}\left(\frac{kN+x}{2\sqrt{Dt}}\right)+ {\rm erfc}\left(\frac{N(1+k)-x}{2\sqrt{Dt}}\right)\right\}
\right]\nonumber\\
&-&\frac{(1-\rho_0)e^{-Jt}}{\frac{{\cal T}}{2}\sqrt{\pi t/D}}\sum_{k=0}^{\infty}(-1)^{k} \left\{
\exp{\left[-\frac{(kN+x)^2}{4Dt}\right]}+ \exp{\left[-\frac{(N[1+k]-x)^2}{4Dt}\right]}
\right\}.\nonumber\\
\end{eqnarray}
As for an infinite chain, it follows from Eqs.(\ref{clongper1}) and (\ref{clongper2}) that also on a (finite) periodic lattice the concentration approaches its stationary value exponentially fast with a subdominant power-law prefactor which depends on the initial state of the system: $c_x(t)-c_x(\infty)\sim e^{-Jt}t^{-3/2}$ starting from an intially completely occupied lattice, while $c_x(t)-c_x(\infty)\sim e^{-Jt}t^{-1/2}$ when the initial concentration is 
$0\leq \rho_0<1$. As a difference with the infinite chain (Section 4), one notices that for a finite ring there are an infinite number of subdominant terms in Eqs (\ref{clongper1}) and (\ref{clongper2}).
In fact, one can also verify that in the thermodynamic limit ($N\to \infty$), the expressions (\ref{clongper1}) and (\ref{clongper2}) simplify considerably since only the term with $k=0$ contributes. In that case, as expected, 
one recovers the results (\ref{nx1d}) and (\ref{nx1CII}) obtained for an infinite chain.

Clearly, one could also consider the effect of periodic boundary conditions in 2D (the lattice would be a torus) and 3D. However, as already illustrated by the one-dimensional situation, the mathematical treatment becomes somewhat involved and the final results deviate from those obtained in Sections 4-7 only for finite size systems.


\vspace{0.55cm}
 {\bf References}
 \vspace{0.25cm}
 
\begin{thebibliography}{99}
\bibitem{Privman}
{\it Nonequilibrium Statistical Mechanics in One Dimension} edited by V. Privman  (Cambridge University Press, Cambridge, 1997);
S. Redner, {\it A guide to first-passage processes} (Cambridge University
Press, Cambridge, 2001);
F. C. Alcaraz, M. Droz, M. Henkel and V. Rittenberg, Ann. Phys. (N.Y.) {\bf
  230}, 250 (1994);
 M. Henkel, E. Orlandini, and J. Santos, {\it ibid} {\bf 259}, 163 (1997);
 D. C. Mattis and M. L. Glasser,
Rev. Mod. Phys. {\bf 70}, 979 (1998);
H. Hinrichsen, Adv. Phys. {\bf 49}, 815 (2000);
U. C. T\"auber, M. Howard and B. P. Vollmayr-Lee, e-print: cond-mat/0501678;
B. Schmittmann abd R. K. P. Zia, in {\it Phase Transitions and Critical}, edited
by C. Domb and J. L. Lebowitz (Academic Press, New York, 1995), Vol. 17.
%

\bibitem{Schutzrev}
G. M. Sch\"utz, in {\it Phase Transitions and Critical Phenomena}, edited
by C. Domb and J. L. Lebowitz (Academic Press, London, 2000), Vol. 19.
%

\bibitem{DPAC}
Z. R\`acz, Phys. Rev. Lett. {\bf 55}, 1707 (1985); 
A. A. Lushnikov, Zh. Eksp. Teor. Fiz. {\bf 91}, 1376 (1986) [Sov. Phys. JETP
{\bf 64}, 811 (1986)];
 M. D. Grynberg, T. J. Newman and R.B. Stinchcombe Phys. Rev. E {\bf 50}, 957
 (1994);
M. D. Grynberg and R.B. Stinchcombe, {\it ibid} {\bf 52}, 6013 (1995);
M. D. Grynberg and R. B. Stinchcombe Phys. Rev. Lett. {\bf 74}, 1242 (1995); 
 M. D. Grynberg and R.B. Stinchcombe, {\it ibid} {\bf 76}, 851 (1996);
P.-A. Bares and M. Mobilia, Phys. Rev E, {\bf 59}, 1996 (1999);
 M. Mobilia and P.-A. Bares,  {\it ibid}, {\bf 63}, 056112 (2001);
M. Mobilia,  {\it ibid}, {\bf 65}, 046127 (2002);
M. Mobilia, B. Schmittmann and R. K. P. Zia, to appear in Phys. Rev. E, {\it at press} (e-print: cond-mat/0412576)
.
%
\bibitem{TMMC}
N. Kuroda, M. Nishida, Y Tabata, Y Wakabayashi, and K. Sasaki, Phys. Rev. B
{\bf 61}, 11217 (2000); N. Kuroda, Y. Tabata, M. Nishida and M. Yamashita,
{\it ibid} {\bf 59}, 12973 (1999); 
N. Kuroda, Y. Wakabayashi, M. Nishida, N. Wakabayashi, M. Yamashita and
N. Matsushita, Phys. Rev. Lett. {\bf 79}, 2510 (1997).
%
\bibitem{MX}
R. Kroon, H. Fleurent and R. Sprik, Phys. Rev. E {\bf 47}, 2462 (1993);
R. Kopelman, C. S. Li and R. Sprik, J. Lumin, {\bf 45}, 40 (1990). 
%
%

\bibitem{Phototrap}
S. H. Park, H. Peng, S. Parus, H. Taitelbaum and R. Kopelman,
 J. Phys. Chem. A {\bf 106}, 7586 (2002).
%
\bibitem{Photobleach2D}
S. H. Park, H. Peng, R. Kopelman, P. Argyrakis and H. Taitelbaum,
 Phys. Rev. E {\bf 67}, 060103(R) (2003).
%

\bibitem{Rittenberg}
H. Hinrichsen, V. Rittenberg and H. Simon, J. Stat. Phys. {\bf 86}, 1203 (1997).
%
\bibitem{benA}
D. ben-Avraham, Phys. Rev. E {\bf 58}, 4351 (1998);
D. ben-Avraham. Phys. Lett. A {\bf 249}, 415 (1998).

%
\bibitem{IVM}
M. Mobilia, Phys. Rev. Lett. {\bf 91}, 28701 (2003).
%
\bibitem{MI}
M. Mobilia and I. T. Georgiev, Phys. Rev. E {\bf 71}, 046102 (2005), {\it at press}
(e-print: cond-mat/0412306).

%
\bibitem{Ben}
E. Ben-Naim, S. Redner and G. H. Weiss, J. Stat. Phys. {\bf 71}, 75 (1993) 
%
\bibitem{Weiss}
G. H. Weiss, R. Kopelman and S. Havlin, Phys. Rev. A {\bf 39}, 466 (1989)
%
\bibitem{Taitelbaum2}
 H. Taitelbaum, R. Kopelman, G. H. Weiss, and S. Havlin, 
Phys. Rev. A {\bf 41}, 3116 (1990)
%
\bibitem{Taitelbaum}
 H. Taitelbaum, Phys. Rev. A {\bf 43}, 6592 (1991)
%

%
\bibitem{Abramowitz}
M. Abramowitz and  I. Stegun, {\it Handbook of Mathematical Functions}
(Dover, NY, 1965); I. S. Gradshteyn and I. M. Ryzhik,
{\it Table of Integrals, Series and Products}
 (Academic Press, San Diego, 1963);
{\it Tables of integral transforms}, edited by A. Erd\'elyi (Mc Graw-Hill, NY, 1954).
%
\bibitem{Itz}
C. Itzykson and J.-M. Drouffe, {\it Statistical field theory, Vol 1}
(Cambridge University Press, New-York, 1989). 

%
\bibitem{Carslaw}
H. S. Carslaw and J. C. Jaeger, {\it Conduction of Heat in Solids}, 
2nd ed. (Oxford University Press, Oxford, 1959).
%
\bibitem{Book}
F. W. Byron Jr. and R. W. Fuller, {\it Mathematics of Classical and Quantum
  Physics}, (Dover, New-York, 1992).
%
\bibitem{Delves}
R.  T. Delves  and G. S. Joyce, Ann. Phys. (NY) {\bf 291}, 71 (2001).

\bibitem{Spitzer}
F. Spitzer, Adv. Phys. {\bf 5}, 246 (1970); 
T. M. Liggett, {\it Interacting Particles Systems}, New York, Springer  (1985).
%
\bibitem{Glasser}
M. L. Glasser and J. Boersma, J. Phys. A {\bf 33}, 5017 (2000). 
%
\bibitem{Glauber}
R. J. Glauber, J. Math. Phys. {\bf 4}, 294 (1963).
%

\end{thebibliography}
\end{document}